# Teaching an LLM agent to fit XRR curves with X-Ray Calc 3

Oleksiy V. Penkov[a,b,*], Jingjing Peng[a,c], Haoyu Fu[a,c]

[a] *ZJU-UIUC Institute, Zhejiang University, Haining, Zhejiang 314400, People's Republic of China*

[b] *Department of Mechanical Science and Engineering, University of Illinois Urbana-Champaign, Urbana, IL 61801, USA*

[c] *School of Materials Science & Engineering, Zhejiang University, Hangzhou 310058, China*

[*] Corresponding author. E-mail: openkov@illinois.edu

## Abstract

The structure of a periodic multilayer X-ray mirror is obtained by fitting its X-ray reflectivity (XRR) curve, and the result depends on how the operator normalizes and trims the curve, frees parameters, and accepts a fit. The manual of the fitting program and the papers describing its engine leave these decisions to the operator, whose practice is tacit, so the fitting stays with the expert. To solve this problem, we proposed to develop a skill for a large language model (LLM) agent via elicitation: the expert's decisions were recorded during fitting and written as thirteen steps and a 22-item report template. The agent runs X-Ray Calc 3 through a Model Context Protocol (MCP) tool server. Fresh sessions, each given the skill, one curve, and a nominal design, were scored against fits the expert had withheld, under six tolerances fixed beforehand. The skill was developed on XRR curves of Co/C mirrors and of Ru/C mirrors from a public data deposit. The final version of the skill was tested on W/B4C multilayers. It was demonstrated that the skill recovered the mean period within 0.3 Å of the expert's fits and the period drift through the stack on both W/B4C specimens, and the W and B4C thicknesses within 1 Å on one of them.

**Keywords:** X-ray reflectivity; periodic multilayer X-ray mirror; curve fitting; tacit knowledge; written procedure; reproducibility; large language model agent; Model Context Protocol

## 1. Introduction

A periodic multilayer X-ray mirror must hold its period and its thickness ratio to within a few Å over tens of bilayers. The reflectance of such a stack arises from the interference of the reflections at every interface. The acceptance measurement, X-ray reflectivity (XRR), does not give the structure directly. Instead, the measured curve is fitted through a model whose thicknesses, widths, and densities are free using LFPSO optimization (Penkov et al., 2024; Li et al., 2023). The reported structure is the parameter set the optimizer settles on.

That result rests on decisions of an operator taken before the optimizer runs, on how the curve is conditioned, and on when the fit is accepted. The papers that define the engine specify the simulation, the cost function, and the optimizer, and say nothing about conditioning or acceptance (Penkov et al., 2024; Penkov et al., 2020; Li et al., 2023). Neither does the engine's manual. Every such decision is left to the operator, whose practice is tacit. That XRR results depend on the operator has been shown by interlaboratory round robins on nominally identical specimens (Colombi et al., 2008; Matyi et al., 2008) and by repeated evolutionary refinement of one dataset treated statistically (Windover et al., 2014). In neutron reflectometry, the analysis has been examined for its dependence on the analyst's model (Shiaelis et al., 2024) and for how many parameters a dataset supports (McCluskey et al., 2020). Thus, a fitted structure can be reproduced only as far as the operator's decisions are written down.

What is required is a written fitting procedure, a text that states each conditioning and acceptance decision of an XRR fit as a step with its reason. The procedure handed to an agent is called a skill. Such a procedure has to be recovered from the expert. It is not held in a form the expert can write out (Polanyi, 2009; Nonaka, 1994; Feigenbaum, 1977). Eliciting such a procedure is an old problem with a method literature (Olson & Rueter, 1987; Cullen & Bryman, 1988; Hoffman et al., 1995). Recent work with an agent as the recipient has turned laboratory notes into executable procedures (Liu et al., 2026b), defined and classified the written procedure as an artifact (Jiang et al., 2026), and managed procedural memory across tasks (Belikova et al., 2026).

Whether such a text is complete can be tested only by a reader who has none of the expert's habits and follows the text literally. A large language model (LLM) agent started in a new session, given only the fitting program, the text, and one measured curve, is such a reader. It has no fitting habits, no memory of other curves, and no access to the expert. Because agreement reached in conversation with the expert proves nothing, the test has to be against fits the expert made and withheld. Fits already in the expert's record serve that test at no refitting cost. Agent results are judged against expert ground truth (Chen et al., 2024; Ai et al., 2026), although one run's success rate hides the inconsistency between runs (Rabanser et al., 2026; Yao et al., 2024). Agreement with an expert is not by itself evidence of quality (Liu et al., 2026a). Automated and machine-learning analysis of reflectivity returns parameters and leaves the verdict with the objective function or the operator (Greco et al., 2019; Greco et al., 2022; Munteanu et al., 2024; Pithan et al., 2023; Sivia & Webster, 1998; Wormington et al., 1999; Björck & Andersson, 2007;

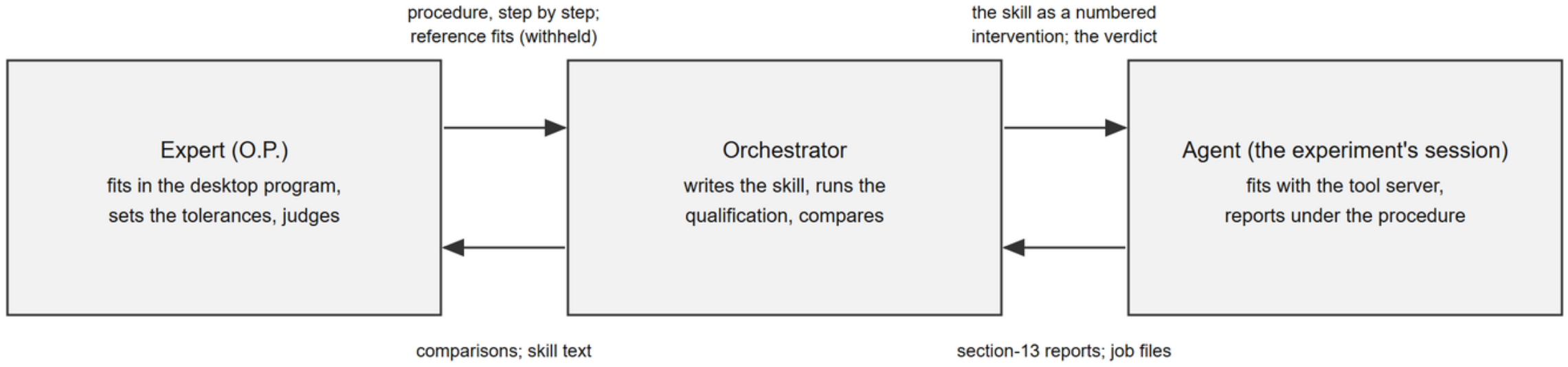

*Figure 1. The expert, the orchestrator, and the agent, and what crosses each boundary. The agent never sees the expert's reference fits or the tolerances, the expert never edits the written procedure, and the fitting server is shared by all three.*

Politsch & Cevc, 2002). Agents have been applied to existing scientific data in powder diffraction, in structure refinement, and in infrared spectroscopy following an expert's own procedure (Wu & Sun, 2026; Li et al., 2026; Shin et al., 2026; Noh et al., 2025). To our knowledge, no published work elicits an expert's own conditioning decisions, writes them down, and verifies an agent's fits against that expert's withheld results.

In this study, an expert's fitting procedure was elicited at the keyboard, written down, and tested blind by LLM agent sessions on Co/C, Ru/C, and W/B4C multilayers. The written conditioning steps reproduced the expert's layer thicknesses within 1 Å on Co/C and Ru/C, as well as the period and its drift through the stack on W/B4C. The acceptance rules, calibrated on one curve shape, did not carry over. Figure 1 shows the expert, the orchestrator, and the agent, and what crosses each boundary.

## 2. System

### 2.1 Specimens and measurement

The specimens are four Co/C periodic multilayers, CoC2 to CoC5, one Co film, Co1, and one single carbon film on glass, C1; five Ru/C periodic multilayers, RuC1 to RuC5, and one Ru film, Ru1, from a public data deposit (Penkov et al., 2026a); and two W/B4C periodic multilayers, WB1 and WB2. Section S2.1 describes their deposition. The XRR curves were measured using a commercial X-ray diffractometer with a W/Si Göbel mirror (PANalytical Empyrean) in a 2θ–ω scan with Cu Kα radiation of 1.54187 Å.

### 2.2 The fitting engine and the tool server

Fits are computed by X-Ray Calc 3 (Penkov, 2026), accessed as a tool server over the Model Context Protocol (Model Context Protocol, 2026). The optical model, the cost function, and the optimizer, a Lévy-flight particle swarm (LFPSO), are as the group's own papers describe them (Penkov et al., 2024; Penkov et al., 2020; Li et al., 2023). The cost function gained one term during the study and is therefore given here in full:

$$\chi^2 = \frac{1000}{N-1}\sum_i w_{p,i} w_{\theta,i}\left(\frac{\log_{10} I_{meas,i} + a - \log_{10} R_{calc,i}}{\log_{10} R_{calc,i}}\right)^2 \quad (1)$$

Here, $I_{meas}$ is the measured intensity at angle θ, $R_{calc}$ is the calculated reflectivity, and $N$ is the number of points. The peak weight $w_p$ is $I_{meas}$ divided by its moving average where that ratio exceeds 3, and 1 elsewhere, so that the Bragg peaks count more than the background. The angle weight $w_\theta$ is $\theta^2$ under $\theta^2$ weighting, with θ in degrees, and 1 without it. The new term is the offset $a$, which rescales the measured curve within a set window at every evaluation. It was zero in campaigns 1 and 2, where the scale was set before the fit, and was solved in campaign 3.

The server exposes calls to compute a model curve, submit a fit against a measured curve, track it, and read back a measured curve. The conditioning of a fit was set per call by the caller. Normalization was passed as a number the caller computed for itself. Five changes were made to this surface during the study, and Section 7 reports them.

## 2.3 Changes in software since the published description

Penkov et al. (2024) describe a graphical desktop program that computes on the CPU, one segment of the angular range per core. It is driven through that interface alone. Two later changes matter here: fitting on a graphics processor (GPU) and a tool server through which an agent calls the program.

Since version 3.9.0, a Direct3D 11 compute shader evaluates the whole LFPSO population in one pass, with no extra runtime on Windows 10 or later. The pass covers the reflectivity of every particle at every angle, the resolution convolution, and $\chi^2$. The CPU still builds the models and rescores the best particle, so the reported $\chi^2$ and curve are the CPU engine's. Wavelength scans and any GPU failure fall back to the CPU. Per iteration, an RTX 5080 was faster than 32 threads of a Ryzen 9 9950X3D by a factor of 2 on a 10-period Si/Mo model of 200 points. For a 150-period W/$B_4C$ model of 2,000 points, it was 73 at a population of 500 and 173 at 5,000. Every fit and every $\chi^2$ reported here was computed or recomputed with the GPU. Each fit record names the device.

XRC_MCP, the tool server added in version 3.8.0, is a console executable that speaks JSON-RPC 2.0 over standard input and output. It runs the graphical program's calculation and fitting code, with the same $\chi^2$ definition. Projects it saves open in that program. Each session is confined to its own directory, with a log of every call. Fits are seeded and reproducible.

## 2.4 Roles

Four parties act in this study. Figure 1 shows the first three. The expert (O.P.) fits curves in the desktop GUI. The expert's verdict decides whether a fit is accepted. The orchestrator is the Claude Code session, model Claude Fable 5.1 (`claude-fable-5-1`), that recorded the expert's fitting procedure.

The agent is an autonomous session, model Claude Opus 5 (`claude-opus-5`), which fits the laboratory's curves. Every message reaching it from outside is a numbered intervention approved beforehand. The test sessions are fresh qualification sessions of model Claude Sonnet 5 (`claude-sonnet-5`). Each received the written procedure and one measured curve, with no memory of any other session.

## 3. What the procedure specifies

Table 1 gives the nine conditioning decisions. The documentation leaves most of them unspecified and is inconsistent with the expert's practice on the one it specifies. For normalization, the manual, the tool description, and the expert name three different targets (S2.3.1).

**Table 1.** The nine conditioning decisions of an XRR fit that no documentation settles, the expert's choice for each, and whether the choice held when tested against a known structure.

| Decision | What the documentation gives | The expert's choice | Step (Table 2) | Tested against known truth (Section 7.6) |
|---|---|---|---|---|
| Normalization point | Manual: comparison at about 0.4°. Tool description: the total-reflection plateau | The plateau maximum | 4 | revised: the scale is solved inside the objective |
| Smoothing | Nothing | One pass, always | 5 | revised: off |
| Right limit of the fitting range | Nothing | Past the last visible order | 6 | revised: at least the first seven predicted orders |
| Background floor | Nothing | Read off the curve in hand | 6 | not tested |
| Parameter bounds | Nothing | Generous, because the optimizer misbehaves near a bound; density capped at bulk | 7 | held |
| Angular resolution $\Delta\theta$ | No beam-divergence figure for the instrument | Tuned from the contrast of the secondary fringes, 0.012 to 0.015° here | 8 | held |
| Optimizer settings | An empty heading in the manual | Population size before iteration count | 8 | held |
| Priority when $\chi^2$ and the orders disagree | Nothing | The order peaks before the fringes | 10 | not tested |
| Fitted density | Nothing on its status | Free; a low value read as intermixing | 11 | held |

A $\Delta\theta$ chosen by comparing $\chi^2$ (Eq. 1) alone reached 0.025°, against the 0.012 to 0.015° interval in use in this laboratory (S2.3.2). The order peaks are judged before the fringes, because $\chi^2$ alone can be driven low by a model that misses them. Seven of the nine were tested against curves computed from a known structure, and the choices that failed the test were replaced (Section 7.6).

## 4. Eliciting and writing the procedure

The procedure was elicited by think-aloud observation while the expert fitted specimen C1, a single carbon film on glass, in the desktop program. The expert typed each step and its reason into the orchestrator session while fitting. The orchestrator recorded each step against its tool-server equivalent.

Under a one-layer model, the carbon film precisely yielded only its thickness; the roughness and density are estimations. The one-layer model shows a shoulder at the critical edge that the data lack. The periodic case came instead from the expert's saved project for specimen CoC5, with a thickness profile across the twenty periods chosen from the symmetry of the high-order peaks.

The session was condensed into thirteen numbered steps, each with the reason for it (Table 2). The procedure ends in a 22-item report template, reproduced in the Supplementary Information. Every item has to be filled, so a fit cannot be judged from a number the agent chose to show. The procedure is published as its own record (Penkov et al., 2026b), with the four versions the sessions received. The agent holds it as a skill file, the form in which a session loads a written procedure. Constants of this laboratory were separated from the general rules, so that a reader at another instrument can replace them without rewriting the procedure: the substrate roughness of 3.8 Å, the wavelength of 1.5406 Å, and the Δθ interval of 0.012 to 0.015°. Three exclusions were deliberate: the expert's own fitted numbers, so that the agent could later be qualified on unseen fits; any material system beyond Co/C, so that the transfer test would exercise a procedure written for one system only; and any narrative of the session, so that the procedure reads as instructions.

The written procedure was audited against 51 of the expert's own project files, to test it against the expert's recorded practice rather than the elicited account. Δθ ranged from 0.009 to 0.026° and was set by measurement in each case rather than fixed by the instrument. The LFPSO swarm population, the number of candidate structures evaluated in each optimizer iteration, stood at 5000 in every multilayer fit of 2026 except the two CoC4 reference projects, at 500. A thickness profile through the stack was the default in every periodic fit except those same two. In this polynomial mode of order 3, each layer thickness follows a cubic polynomial in the period number, while σ and ρ keep one value per layer. The written procedure agreed with the practice in all three respects.

**Table 2.** The thirteen steps of the written procedure against the expert's GUI session.

| Step (procedure section) | GUI action | Tool parameter | Reason |
|---|---|---|---|
| 1. Read the curve | Copy curve into project; read θ_max/I_max by eye | `get_measurement` (`max_points` 2000) | Anchors normalization (§4) and fitting range (§6) |
| 2. Starting model | Substrate glass, modeled as $SiO_2$, 2.65 g/cm³, σ 3.8 Å, fixed | Substrate fixed; `polarization` "s"; `lambda` 1.5406; surface-layer stack N 1 (added after the qualification, Section 7.4) | 3.8 Å is fit-derived; s-polarization halves run time |
| 3. Sensitivity check | No GUI equivalent | `calc_reflectivity` on one-parameter variants | Maps which curve feature each parameter moves |
| 4. Normalize (scale) | Data → Normalize Auto | `scale: "auto"`, `auto_theta_max` 0.5 | Sets measured maximum equal to model |
| 5. Smooth | Data → Smooth, one pass | `smooth {"passes": 1}` | One pass always; GUI has no settings |
| 6. Low limit and fitting range | Data → Trim; set left/right limits | `theta_range`, `r_min` | Right limit past last order; r_min at background |

| 7. Free parameters and bounds | Fitting table, all parameters free, wide bounds | `free`, `bounds` | Optimizer misbehaves near bounds; density capped at bulk |
|---|---|---|---|
| 8. Settings | Calculation-settings pane, set N, θ, λ, Δθ | `resolution`, `chi2`, `optimizer` | Δθ tuned from fringe contrast; population before iterations |
| 9. Periodic or profile mode | Mode Polynomial vs. Periodic, by peak symmetry | `profile`, `poly_order` 3, `paired` | Asymmetric high orders show thickness drift through stack |
| 10. Judge the fit | Watch chi-squared (KPI) during and after fit | `report` (orders, edge, fringes, bands, near_bounds) | $Chi^2$ alone is insufficient; peaks judged before fringes |
| 11. Density | Not a GUI action; read off fitted value | Stays in `free`/`bounds` only, never judged | Free parameter only; a low value signals intermixing |
| 12. Single films | Left limit moved to 2θ 0.6° | `theta_range.min` exception | One-layer model misses carbon-film edge and fringe contrast |
| 13. Report | No GUI equivalent | 22-item report template | Prevents the agent omitting or hiding a number |

## 5. Experimental design

### 5.1 Unit of analysis and endpoints

The unit of analysis is the test session, a fresh model session given the written procedure and one curve, run to a verdict or a stop. A curve fitted in two rounds contributes two sessions. Figure 2 shows the flow of sessions through the three campaigns. The primary endpoint is agreement with the expert's reference fit of the same curve within the tolerances of Table 3, judged by the expert. Each session also recorded its own verdict under the procedure's criteria, without sight of the reference. The two verdicts differ where a fit met every criterion of the procedure and still missed a tolerance, as in round 3 of CoC5 (Section 7.1). The secondary measures are the $\chi^2$ of the accepted fit, the turns, the context tokens, and the wall time. Because the expert's files store the weighting settings and no $\chi^2$, every reference-fit $\chi^2$ was computed after the fact, on the expert's own conditioning and on each session's, by the procedure of S4.1. The comparison favors the expert by an amount that was not quantified, because the reference curve was optimized on the points it is scored on.

### 5.2 The tolerances

The expert fixed six tolerances before any qualification session ran, against two of the expert's own fits. Table 3 gives them. They were applied unchanged to all three campaigns, apart from the roughness tolerance (Section 5.6).

**Table 3.** The six tolerances, fixed before any qualification session ran.

| Quantity | Tolerance |
|---|---|
| Period | within 0.3 Å of the reference |
| Each layer thickness | within 1 Å of the reference |
| Each roughness (σ) | within 1 Å of the reference, on the correct interface |
| Every Bragg order visible above the background | calculated / measured peak intensity between 0.75 and 1.25 |
| $\chi^2$ | under 10 |
| Density | not judged |

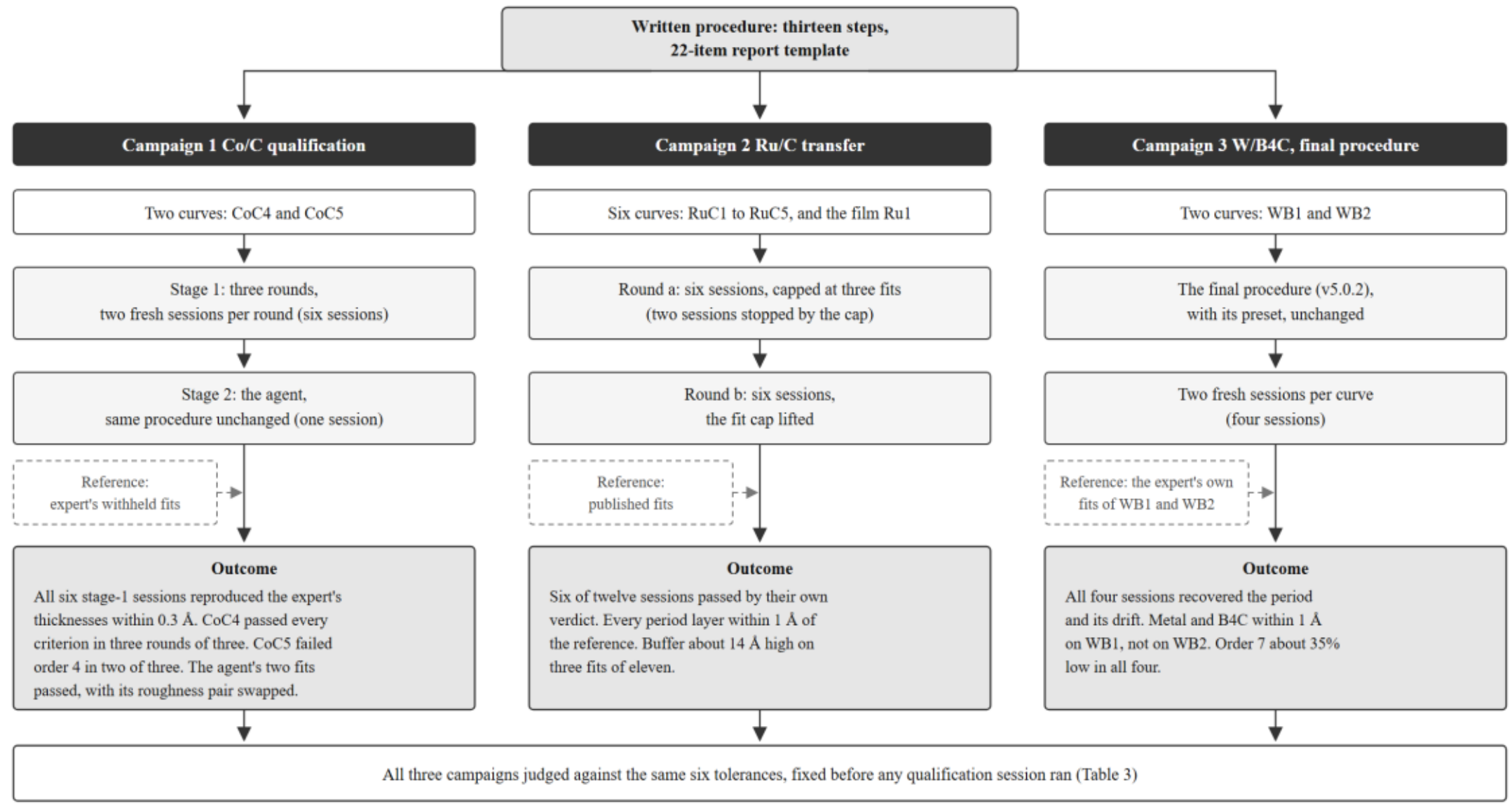

*Figure 2. Flow of sessions through the three campaigns. Campaign 1 qualified the written procedure on CoC4 and CoC5 in three rounds of two fresh sessions, and then gave it unchanged to the agent. Campaign 2 applied the same artifact to six Ru/C curves in two rounds of six sessions. Campaign 3 applied the final procedure to two W/B4C curves, two fresh sessions each. Dashed boxes give the reference each session was judged against.*

The 25% band on the orders is tighter than the agreement the expert's own fits reach. The two reference fits sit between 0.69 and 1.76 on the orders, so several of their orders fall outside the band set for the sessions. The expert kept it (S3).

## 5.3 Blinding and provisioning

Each session received the fitting procedure, the measured curve, and a nominal design, and no fitted number. Sessions had no file or web access and reached the fitting engine only through the tool server. The expert's Co/C reference fits were withheld. The expert's Ru/C reference fits are in a public data deposit (Penkov et al., 2026a), but no session could reach them. The Ru/C comparison is blind on the same basis as the Co/C comparison, by what the sessions could reach rather than by the reference being unavailable.

## 5.4 Campaign 1, Co/C qualification

Campaign 1 used two Co/C curves, CoC4 and CoC5, periodic mirrors of different periods, against two withheld fits by the expert. These were the only two curves of the campaign the expert had fitted before any session ran. Stage 1 ran three rounds of two fresh test sessions, one per curve, six in all. Stage 2 gave the same procedure, unchanged, to the agent and had it fit the same two curves. The agent is the session that produced the rejected fits of Section 6.

## 5.5 Campaign 2, Ru/C transfer

Campaign 2 used six deposited Ru/C curves (Penkov et al., 2026a), five multilayers and one Ru film. Twelve fresh sessions ran, two per curve, in rounds a and b. The procedure was amended once after qualification, as Section 5.6 records, and then applied without further change, including the density bounds and start values written for Co/C.

Each session received the curve and the nominal design: a period of about 68.5 Å, Ru about 15 Å, C about 54 Å, and a Ru adhesion layer of about 200 Å nominal. Three curves were included because their realized structure had departed from that design. RuC2 and RuC3 were realized with a period about 20% above nominal, and the film Ru1 about 35% thinner. On those three, the structure had to be recovered from the curve rather than from the design.

## 5.6 Protocol changes during the study, declared

Six changes were made while the study ran. First, the procedure text was revised three times: between rounds 1 and 2, between rounds 2 and 3, and after qualification, before the transfer. The first revision stated the density bounds explicitly and narrowed the band a session applied to its own orders, from a factor of two to 25%. Between rounds 2 and 3, the judgment of a fit moved onto the report the server had gained. Before the transfer, a carbon surface layer entered the starting model. Section S1 gives the transfer text in full, the differences between all five texts, and their hashes. Only the two round-1 sessions ran without the density bounds and on the wider band.

Second, the rule governing when a profile fit may replace a periodic fit was rewritten and renumbered, from step 9.3 to step 9.4. The earlier text kept whichever of the two fits had the lower $\chi^2$. A passing periodic fit is now final. A profile fit is kept only where the periodic fit failed and passes every criterion at a lower $\chi^2$. Only the six stage-1 sessions ran under the earlier text.

Third, the agent-facing surface of the tool server was revised twice, once between rounds 2 and 3 and once before the Ru/C sessions. Section S2 records which build each session ran on. Fourth, the cap of three fits per session, which stopped two sessions of round a, was lifted for round b. The expert ruled it normal practice (S3). Fifth, the roughness tolerance of Table 3 was dropped in stage 2 of campaign 1. Because the optimizer's seed decides the roughness assignment within a period (Section 7.4), the expert ruled that it is not judged when the peaks pass (S3). From then on, layer thicknesses were scored, and roughnesses were not.

Sixth, campaign 3 ran the final procedure, which carries every laboratory constant in a separate preset file. Compared with campaign 2, it revises the normalization, the smoothing, and the fitting range as Table 1 marks, sets the $\chi^2$ acceptance limit at 1.3 under the solved scale, and reads the instrument's raw file (S1.1). $\theta^2$ weighting is turned off, the wavelength is 1.54187 Å instead of 1.5406 Å, five optimizer seeds are run instead of three, and interlayer thicknesses are held constant through the stack in profile fits. The first three changes follow from the synthetic benchmark (Section 7.6; S5). Without $\theta^2$ weighting, the fitted Co thickness on Co/C came closer to the truth in 80% of noise realizations, at an rms error of 0.05 Å against 0.13 Å. Only the period favored the weighting, by 0.01 Å. The measured curves carry K$\alpha$2, so the fitted period came out 0.055 Å short at 1.5406 Å, the K$\alpha$1 wavelength. At 1.54187 Å, the intensity-weighted mean of the doublet, it came within 0.014 Å. Three seeds understated the spread between seeds about 2.5-fold. The spread was stable from five seeds on. The interlayer thicknesses are held constant because the substrate holder is water-cooled, so the deposition rates drift through a stack and the intermixing at the interfaces does not. No two rounds of campaign 1 ran under the same conditions, and round b of campaign 2 did not run under those of round a. Differences between rounds are not evidence of session-to-session variance.

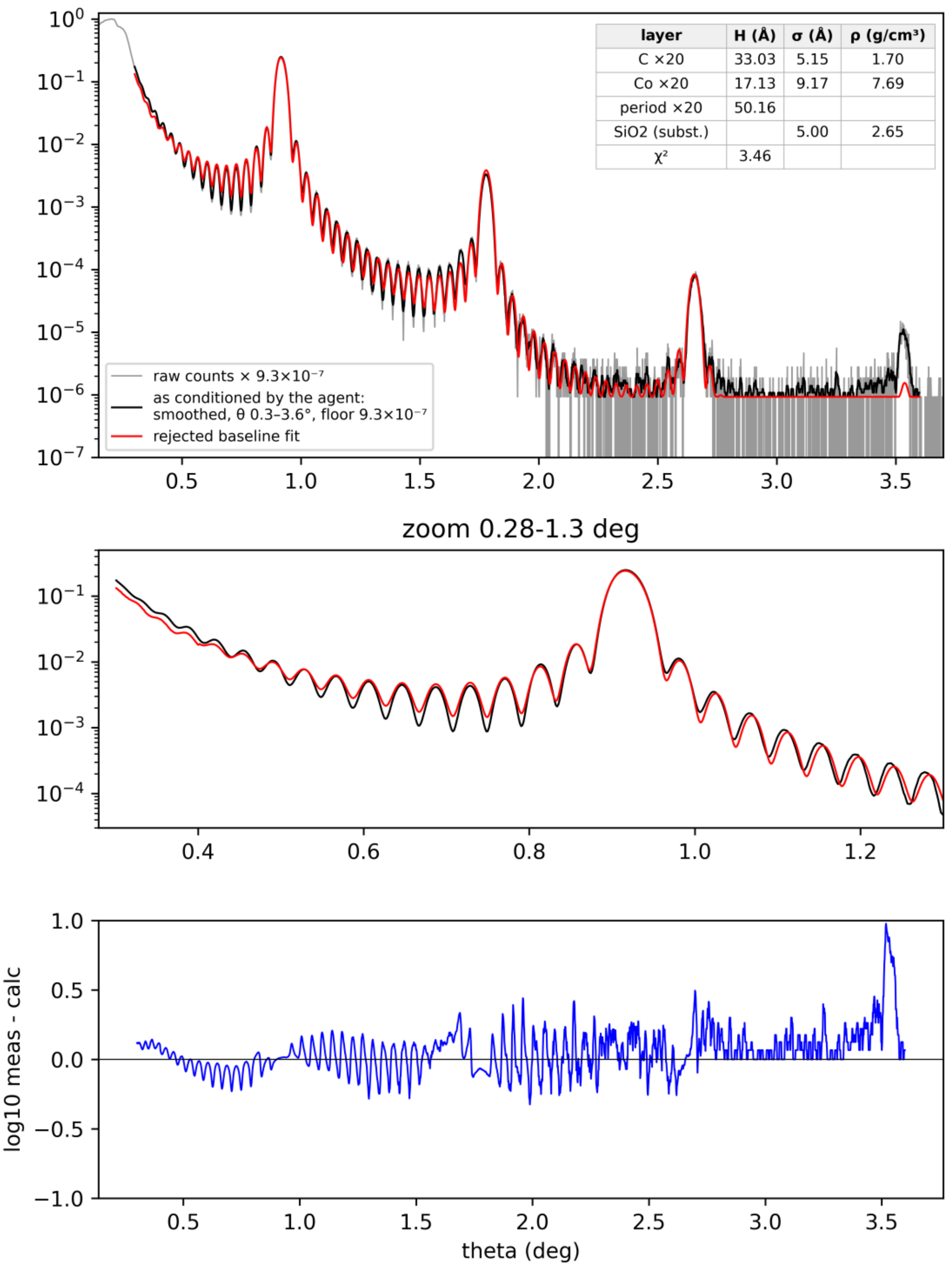

Figure 3. A rejected baseline fit of CoC5 against the measured curve, before the written procedure existed. The fourth Bragg order is about 7 times too low and the model sits on its floor above 2.7 degrees. The table gives the structure of that fit and its $\chi^2$.

### 5.7 Budget accounting

Four quantities are reported per session: turns, tool calls, context tokens, and wall time. Context tokens are the sum of cache-read, cache-creation, and input tokens. Wall time is given in minutes.

### 5.8 Campaign 3, the final procedure on W/B4C

Campaign 3 tested whether the final procedure, run unchanged, recovers the structure of a measured third material system (S6). It ran the final procedure and its preset on X-Ray Calc 3.9.3. Two fresh sessions fitted each of two W/B4C multilayers of 30 periods on glass, WB1 and WB2, from the raw instrument files. The nominal design, the deposition recipe at the laboratory's calibrated rates of 2025, had a period of about 48 Å, W about 18 Å, B4C about 30 Å, and a B4C buffer and cap of about 87 Å. The realized period of about 56 Å had to be recovered from the curve. The references were the expert's profile fits with WC interlayers. The tolerances of Section 5.2 were fixed before the first session. Thickness was scored on B4C and on the metal of a period, W plus both WC interlayers. The W/WC split was recorded but not scored.

## 6. Baseline: fitting before the procedure was written

Before any written procedure existed, three rounds of agent fits were rejected in turn. The first round fitted Co1 and CoC2 (S2.3.3), the second CoC3 and CoC4 (S2.3.4), and the third CoC5 and the carbon film C1 (S2.3.5). Each round shared the same five defects, which makes them a taxonomy rather than three accidents.

The first defect mode is the normalization point. The round-1 fit put its plateau at 0.87 of the measured maximum, against the model's 0.83 (S2.3.6). The second mode appears at the critical edge. The fit of C1 put the model at 0.127 against 0.073 measured, at a θ of 0.26° (S2.3.7). The third is fringe contrast, which ran in both directions inside a single fit. The three visible fringes of C1 measured 3.5, 6, and 6 against 17, 11, and 1.6 calculated. The CoC5 fit put the fourth order about 7 times too low (Figure 3) (S2.3.8).

The background floor is the fourth. The CoC5 model floor was 9.3 x 10^-7 against a measured background of about 1.5 x 10^-6 above 2.7°. The fifth is a roughness parameter absorbing the misfit. In the CoC5 fit σ_Co reached 9.17 Å on a Co layer of 17.13 Å. No fit was accepted until a fitting procedure had been written and the agent qualified against withheld fits (S2.3.9).

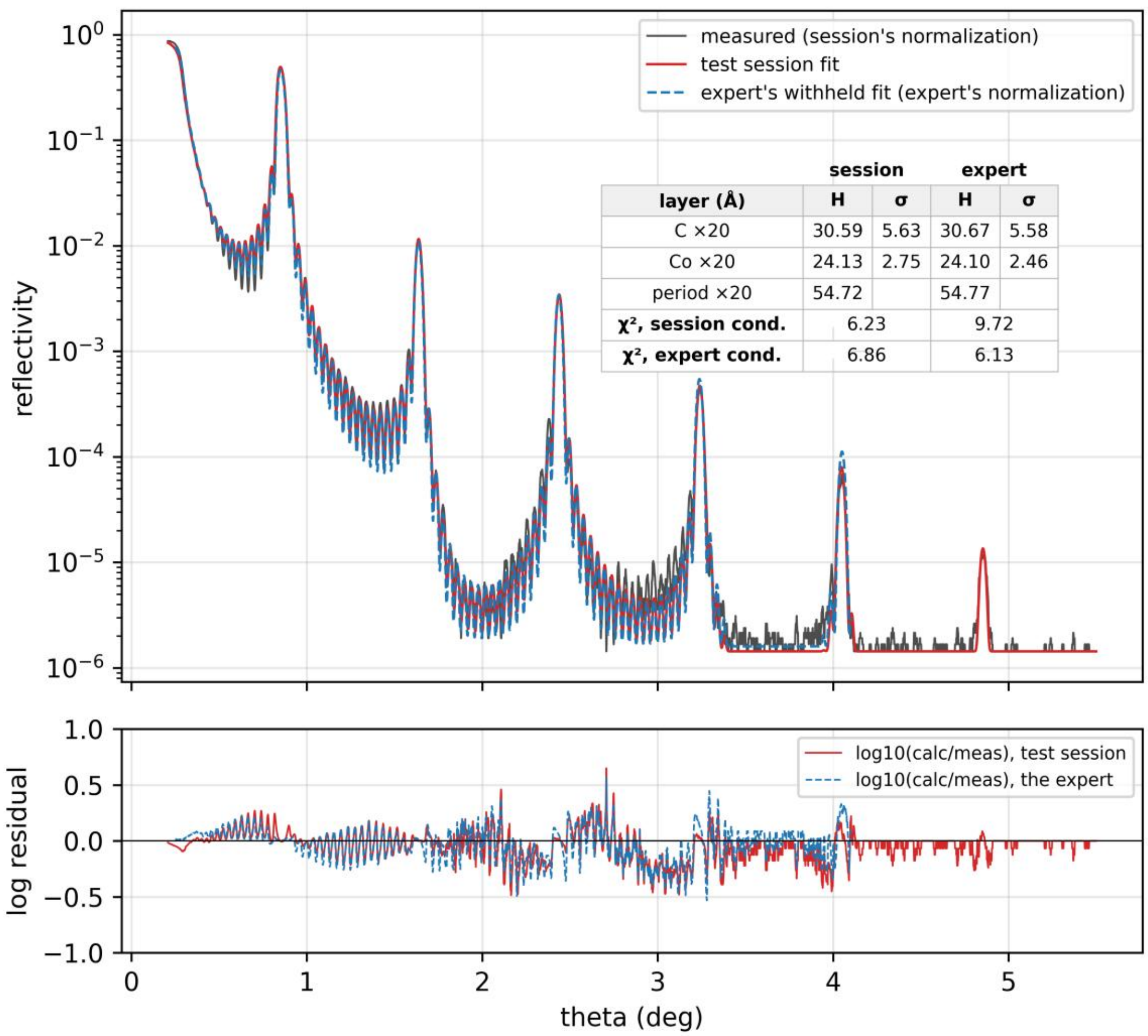

| layer (Å) | session H | session σ | expert H | expert σ |
|---|---|---|---|---|
| C ×20 | 30.59 | 5.63 | 30.67 | 5.58 |
| Co ×20 | 24.13 | 2.75 | 24.10 | 2.46 |
| period ×20 | 54.72 | | 54.77 | |
| **χ², session cond.** | 6.23 | | 9.72 | |
| **χ², expert cond.** | 6.86 | | 6.13 | |

Figure 4. Stage 1, round 1. The fit the test session accepted (red) against the expert's withheld fit (blue dashed), for CoC4. The measured curve is shown on the session's own normalization and the expert's fit on its own. The lower panel gives the log residual of each fit. The table gives the layer thicknesses H and roughnesses σ of the session's fit and the expert's. Each fit's $\chi^2$ is given on the session's conditioning and on the expert's. Values compare only within a row.

# 7. Results

## 7.1 Campaign 1: qualification

Five of the six stage-1 sessions reproduced the expert's withheld thicknesses within 0.3 Å, and the sixth within 0.5 Å. CoC4 passed every criterion in all three rounds. CoC5 failed the 25% band on order 4 in two of three rounds. At the lower $\chi^2$ of 3.51 against 3.60, round 3 of CoC5 accepted a profile fit in which the roughness pair was swapped. Although that fit passed every criterion the session applied to itself, it missed the 1 Å roughness tolerance on both interfaces. Figure 4 overlays the session and expert fits for CoC4. In stage 2, the agent's fit calls were refused as invalid JSON fourteen times (S2.3.10). Given a hand-computed scale from stage 1, it fitted both curves to a verdict in 13 turns (S2.3.11). Against the withheld references, CoC5 passed every tolerance (S2.3.12). CoC4 passed every tolerance but a swapped roughness pair (S2.3.13). The expert ruled that the peaks decide rather than the secondary fringes and passed both curves. Section 7.4 resolves the roughness assignment.

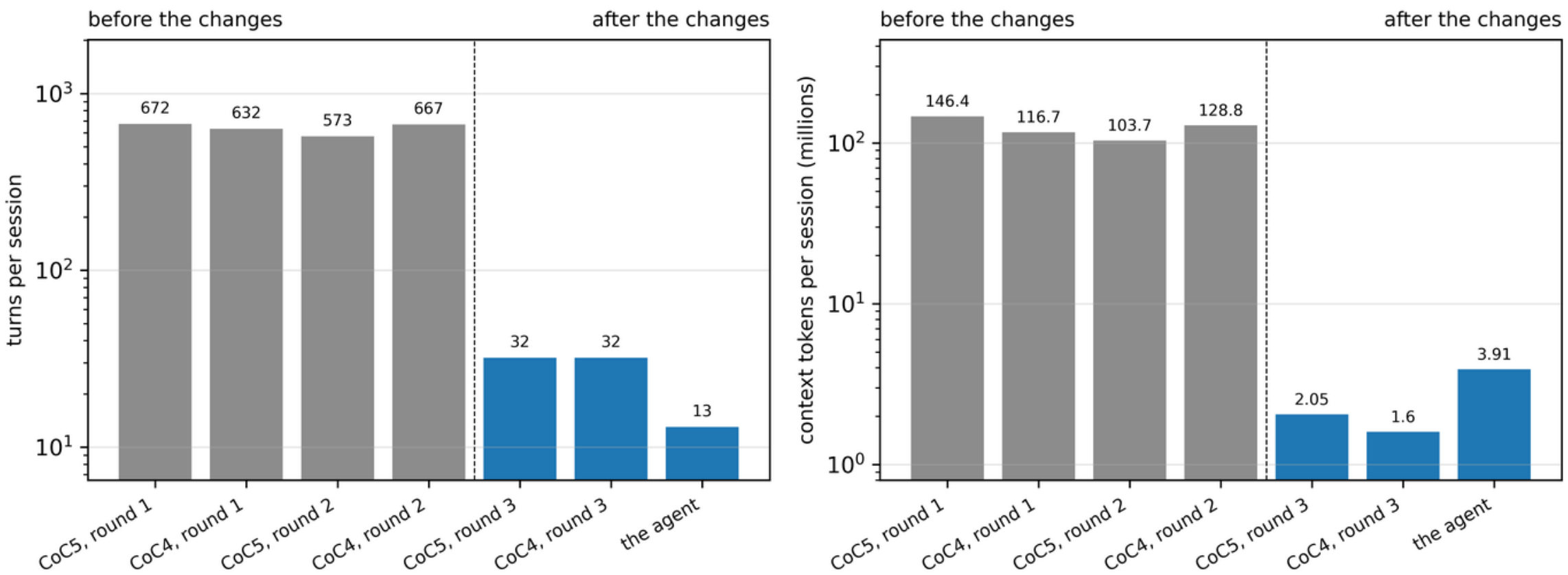

*Figure 5. Cost of a judged two-fit session before and after the five interface changes: turns (left) and context tokens (right), log scale. Grey, the rounds run by polling; blue, the round run with a blocking wait and the per-fit report, and the agent's own session.*

## 7.2 Campaign 2: the Ru/C transfer

Six of the twelve transfer sessions passed by their own verdict. Every period layer of all 18 profile fits, accepted and discarded alike, lay within 1 Å of the expert's. Fourteen of the 18 lay within 0.7 Å. The six fails had three causes: the profile-fit rule of step 9.4 decided four, the fit cap one, and $\chi^2$ alone one. Under the procedure's weighting, the expert's own fitted curves scored 10.17 on RuC1 and 14.59 on RuC4 against the limit of 10. That limit was fixed on Co/C curves of four and six orders and does not carry to curves of ten orders. Section 7.5 gives the mechanism behind the gap. One physical miss went uncaught by the procedure. The Ru buffer reached 144 Å in 3 of the 11 multilayer fits. Because the start at the nominal 200 Å never touched the 133 Å lower bound, the widening rule never fired. The expert's own buffers across the series lie between 13 and 132 Å.

## 7.3 The tool interface and the cost of a session

Stage 1 exposed five gaps in the tool server, all closed before stage 2 (S2.3.14). Each change removed a class of work done by hand: a poll loop of about one turn per second of waiting, about four turns per fit for normalization, and the reading of two 2000-point lists by eye. Figure 5 gives the effect on the cost of a session. A judged two-fit session ran 573 to 672 turns and 104 to 146 million context tokens before the changes. After them, it ran 32 turns and 1.6 to 2.1 million.

## 7.4 The seed decides the roughness assignment

On CoC4, the agent's fit gave σ_C 2.64 Å and σ_Co 4.33 Å at a $\chi^2$ of 7.94 (S2.3.15). The expert's withheld fit gave the opposite pair, 5.58 and 2.46 Å. A stage-1 session found 5.70 and 2.75 Å independently. All three place the same peaks and period and differ only in the fringe field between the orders. Because the swarm is seeded over the whole bound range, two fits started from opposite structures returned identical results (Table 4). A carbon layer on top lowered $\chi^2$ from 7.94 to 6.09 and moved the mean log residual of the band between orders 2 and 3 from +0.133 to −0.064. Nothing in the period stack moved. Thus, the assignment follows the seed and was removed from the tolerances (Section 5.6). Where the peaks and the fringes disagree, the peaks are given priority.

**Table 4.** The roughness assignment of CoC4 across three seeds, without and with a surface layer.

| Seed | Without surface layer: σ_C / σ_Co (Å), $\chi^2$ | With surface layer: σ_C / σ_Co (Å), $\chi^2$; layer thickness, density (g/cm³) |
|---|---|---|
| The agent's | 2.64 / 4.33, $\chi^2$ 7.94 | 2.63 / 4.46, $\chi^2$ 6.09; 12.6 Å, ρ 0.68 |
| Seed 1 | 5.71 / 2.75, $\chi^2$ 7.13 | 4.46 / 2.70, $\chi^2$ 6.23; 13.3 Å, ρ 0.80 |
| Seed 2 | 5.70 / 2.75, $\chi^2$ 7.13 | 2.63 / 4.46, $\chi^2$ 6.09; 12.6 Å, ρ 0.68 |

## 7.5 Fit quality on one footing

Every $\chi^2$ above was computed on its own session's conditioning and does not compare across sessions. The like-for-like procedure of Section 5.1 rescores them in both directions. On the expert's own conditioning, the written procedure halved $\chi^2$ on CoC5, from 6.95 to 3.44, and left the test session 12% above the expert. On CoC4, it cut $\chi^2$ by 17%, to within 7% of the expert. The agent's stage-2 CoC4 fit, with the swapped pair of Section 7.4, scored 8.89, about 45% above the expert. Scored the other way, the expert's structure reached 11.05 on CoC4 against the stage-1 session's 7.28 and 5.49 against 3.60 on CoC5. The order reverses. Because every fit scores better under its own conditioning, the ranking follows the conditioning rather than the model. Over all 158 fits, the factor between the engine's weighted $\chi^2$ and the plain one ran from 0.61 through a median of 3.13 to 22.31 (S2.3.16). Thus, a fixed limit on the weighted value acts through a multiplier that changes from fit to fit.

## 7.6 The procedure against known truth

Sections 7.1 to 7.5 measure agreement with one operator. Only a curve computed from a known structure makes the error observable. Two synthetic multilayers, Co/C with 20 periods and W/B4C with 30, were fitted about 5,000 times on the published engine, with Poisson noise and Kα2 present. The designs and every number are in S5. The period was recovered. The division of a period between a heavy layer and its carbide was not. The solved scale capped what a wrong normalization anchor did to the fitted densities. Two of the procedure's own choices failed the test and were replaced. Because the written procedure carries every conditioning decision as a parameter, the procedure that fit the measured curves ran the benchmark with the same fitting steps. Its final revision changed only the conditions under which the $\chi^2$ limit was calibrated. The truth is a model of a multilayer rather than a multilayer. Thus, the test bounds the inverse step and says nothing about the forward model.

### 7.7 Campaign 3: the final procedure on measured W/B4C

All four sessions read the period off the Bragg orders rather than the design. Each rejected the periodic model on its order pattern and accepted a linear profile fit, at $\chi^2$ between 0.88 and 1.29 (Figure 6; S6.3). The mean period lay within 0.06 Å of the expert's on WB1 and within 0.22 Å on WB2. From surface to substrate, the period grew by 0.74 Å against the expert's 0.76 Å on WB1, and by 1.16 to 1.20 Å against 1.21 Å on WB2. The two sessions of a curve agreed within 0.06 Å in every mean.

On WB1 the metal and B4C lay within 0.32 Å of the expert's. On WB2 the sessions' metal was 1.2 Å thinner than the expert's and their B4C 1.4 Å thicker. The expert's model of that curve carries 8.2 Å of WC around 5.0 Å of W. The sessions kept a two-layer period because $\chi^2$ already met the procedure's acceptance limit of 1.3, and the procedure adds an interlayer only when a fit fails. The profiles also differ in where the drift sits (Figure 7). On WB1 the expert puts it on W and both sessions on B4C. On WB2 all three fits put it on B4C.

No session met the order tolerance. At the scale each fit was scored on, order 7 fell 35 to 38% short in all four, and on WB1 order 4 stood 35 to 39% high. The expert's fits met the tolerance on every order but one. The server reported the order band against the anchored scale rather than the solved one, so each session judged the band on a scale its fit was not scored on.

This is the finding of Section 7.6 on measured curves. The period and its drift were recovered. The division of the period and the assignment of the drift were not, each on one of the two curves.

## 8. Discussion

The written procedure reproduced the expert's Co/C fits within the tolerances fixed beforehand and, applied unchanged, recovered the Ru/C layer thicknesses within 1 Å. On measured W/B4C, the final version recovered the period and its drift through the stack. The acceptance thresholds failed on both new material systems. Within a period, the roughness assignment followed the optimizer's seed. A tool surface built for an agent cut the cost of a judged session by more than an order of magnitude.

### 8.1 Applying the skill to Ru/C and W/B4C

The procedure's steps for conditioning the curve, setting bounds, and choosing settings and a model produced the expert's structures on a material system the procedure was not written for, including curves whose realized structure departed from the nominal design. On those curves, the structure was recovered from the curve rather than the design. The final version did the same on W/B4C, whose realized period was 17% above the design (Section 7.7).

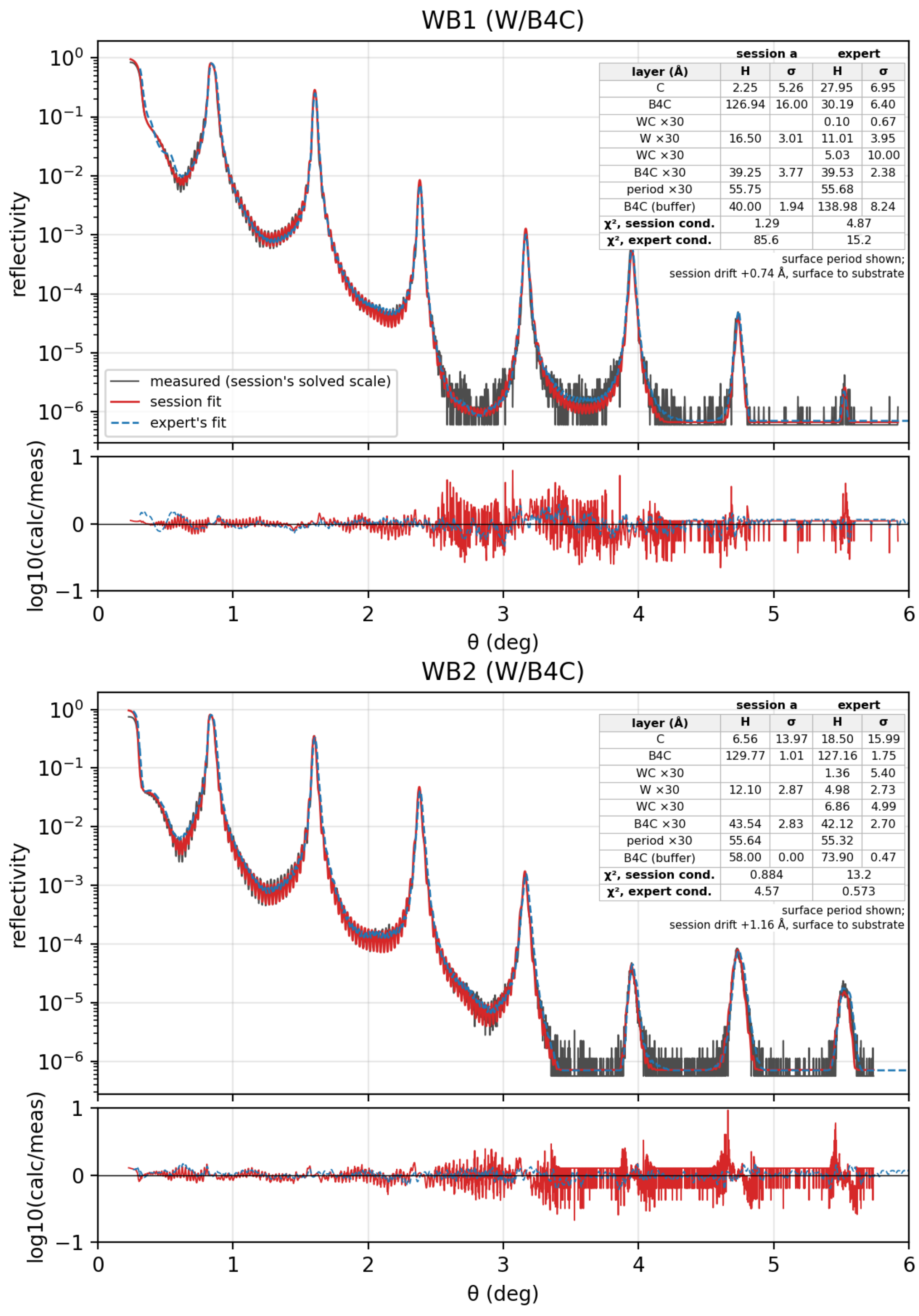

Figure 6. Campaign 3. The fit session a accepted (red) against the expert's fit (blue dashed), for WB1 and WB2. The measured curve is on the session's solved scale, the expert's fit on its own normalization; the lower panels give the log residuals. The tables give the surface period of both fits and each fit's $\chi^2$ on both conditionings, and the note the session's drift of the period through the stack.

The rules of verdict did not carry over from the Co/C qualification curves to Ru/C and W/B4C. The order band and the $\chi^2$ limit were fixed on the qualification curves, which carry four and six orders, and applied unchanged to curves carrying ten. The expert's own fits fail the $\chi^2$ limit the sessions were held to (Section 7.2). On $W/B_4C$ the sessions accepted fits under the $\chi^2$ limit of the final version whose seventh order fell 35 to 38% short, on curves where the expert's fits met the order band on every order but one (Section 7.7). The weighting result of Section 7.5 gives the mechanism, since the same threshold means a different plain $\chi^2$ on every fit. The conditioning steps and the rules of verdict therefore rest on different evidence. A procedure that ships an acceptance threshold has to record the curves it was calibrated on, because a threshold applied outside them reports the calibration rather than the fit.

## 8.2 Agreement with one expert is not truth

The qualification measures agreement with one operator under one conditioning, against the expert's own fit, six tolerances, and both thresholds. Each fit scored better on the conditioning it was made under. Neither fit is better on the evidence available. Interlaboratory work in reflectometry has established that results for nominally identical specimens disagree between laboratories (Colombi et al., 2008; Matyi et al., 2008). A single operator's fit is one draw from that distribution rather than a value the technique returns. The same caution has been raised for the evaluation of language models, where human consensus is not ground truth (Liu et al., 2026a). The qualification is evidence that the procedure reproduces this expert's practice, and none about the true structures.

## 8.3 Non-uniqueness of the roughness assignment

A roughness pair reported from one specular curve and one seed is not evidence of which layer carries the larger roughness (Section 7.4). Recovering a multilayer's parameters from a reflectivity curve is an underdetermined inverse problem, because the measurement carries no phase (Munteanu et al., 2024). A study of nonunique solutions in XRR curve fitting identified the mass density and the surface roughness of a layer as parameters that are weakly determined together (Tiilikainen et al., 2007). In a Mo/Si multilayer the exchange of the two interlayer thicknesses within a period returned almost the same reflectivity once the interfacial roughness was comparable to the interlayer thickness (Nayak et al., 2006). Standard reflectometry fitting algorithms were reported to identify one solution where several structures are compatible with the measurement (Starostin et al., 2025). The ordering of the two assignments by $\chi^2$ is not stable, so an automated fitter reports whichever assignment its seed reaches. That a carbon surface layer moved the fringe field and not the period stack (Section 7.4) agrees with the known limits of XRR under surface contamination (Gil & Windover, 2012).

## 8.4 Interface design as a determinant of agent cost

The reduction in session cost came from a change of interface, with the model, the written procedure, and the curves unchanged. Interface design rather than fine-tuning has been argued to drive the success of tool-using agents (Wu et al., 2025). The baseline was a tool surface written for a human caller. A controlled study separates interface misuse from semantic misuse (Sigdel & Baral, 2026). The schema fault was interface misuse.

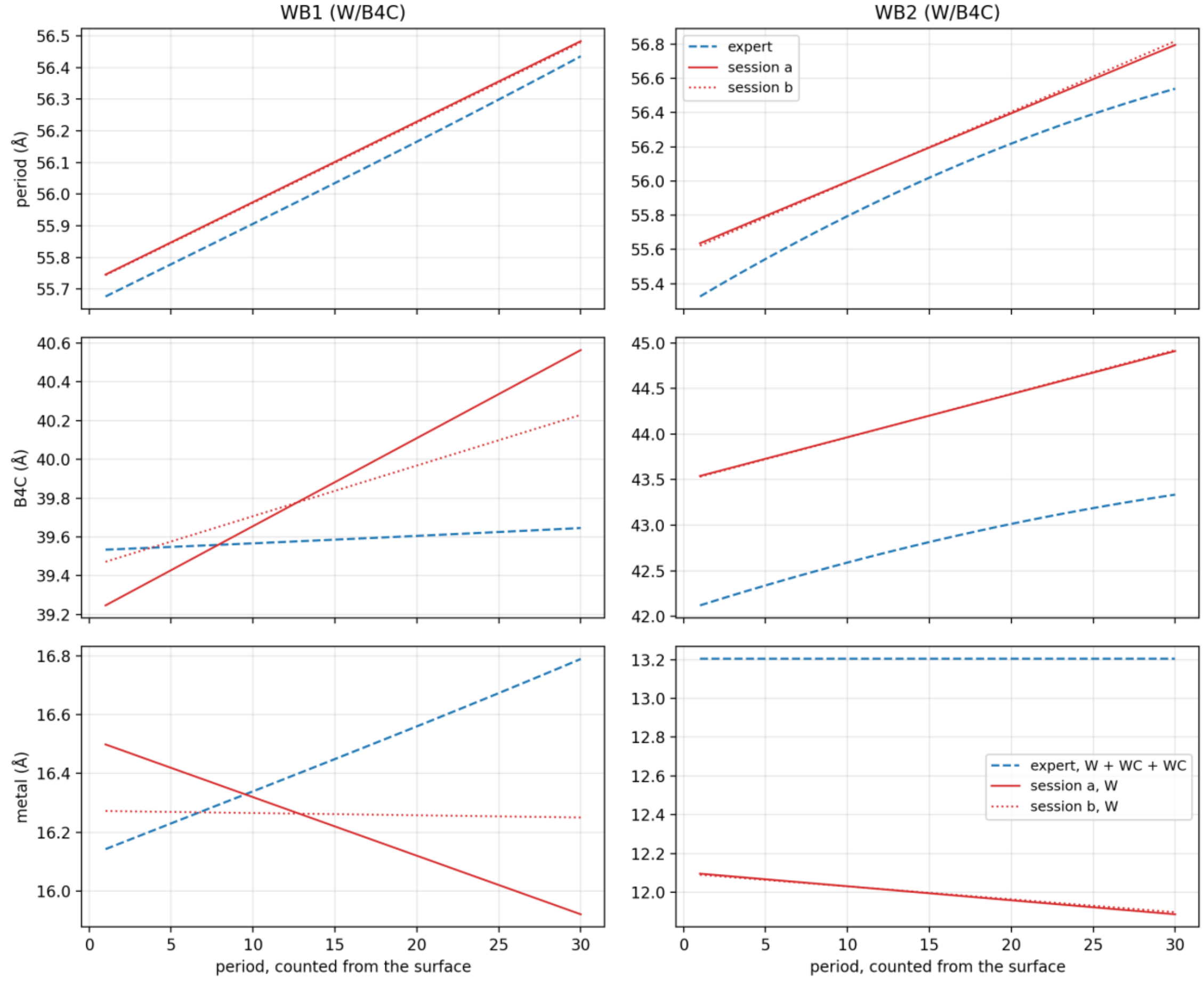

Figure 7. Campaign 3. Period, B4C thickness, and metal thickness of each period, counted from the surface, for WB1 and WB2. Both sessions of each curve (red solid, a; red dotted, b) are shown against the expert's fit (blue dashed). The expert's metal is W plus both WC interlayers. The sessions' metal is W.

## 8.5 Bounds and generalization

No XRR curves were measured on a second diffractometer, so the resolution function, the background, and the alignment are constants rather than variables. A curve from another instrument therefore needs its own resolution and laboratory constants in the preset before the procedure applies (Section 5.6). The procedure has not been tried on a system of low optical contrast or at a synchrotron.

# 9. Conclusions

The decisions of an XRR fit that the engine's documentation leaves open can be written as a procedure that leads an LLM agent to the expert's result without the expert present. The procedure held on Co/C, transferred to Ru/C, and recovered the W/B4C period and its drift through the stack. Its conditioning steps transferred and its thresholds did not, because a $\chi^2$ or order threshold belongs to the curves and the conditioning it was calibrated on. A published procedure has to name those curves. One specular curve decides neither which interface carries the larger roughness nor how a heavy layer and its carbide share a period. An automated fit reports the roughness pair its seed reaches. With the model unchanged, a tool interface rebuilt for the agent cut the cost of a judged session by more than an order of magnitude.

## 10. Program distribution and technical details

The written procedure is free under the Creative Commons Attribution 4.0 licence. Version 1.0 was released on 2026-09-20 in the repository github.com/OleksiyPenkov/xrr-fitting-skill and is archived at doi:10.5281/zenodo.22851595 (Penkov et al., 2026b). It is a Markdown document holding the thirteen numbered steps with their reasons, the 22-item report template, and a block of laboratory constants marked in place.

The fitting engine is X-Ray Calc 3 (Penkov, 2026), the software described by Penkov et al. (2024). It is written in Delphi for Windows and is provided free under the GNU General Public License v3 from https://github.com/OleksiyPenkov/X-RayCalc3/releases as precompiled executables. Its own distribution and build requirements are given by Penkov et al. (2024) and are not repeated here.

The sessions reached the engine only through its Model Context Protocol server component, XRC_MCP (Section 2.3). The server is part of the *X-Ray Calc 3* source tree, in a directory of that name carrying its own README, and is covered by the same license. The Windows installer of version 3.9.3, the release cited here, installs the server beside the two X-Ray Calc executables, the user manual, and the example files. A reader therefore obtains the tool surface the sessions used, either from that installer or by building the source. Running a session requires a client that speaks the Model Context Protocol, the server executable, and a working directory for each session. An example client configuration, and the four server revisions cited here with their identifiers, their dates, and the sessions that ran on each, are in the data deposit.

## Supplementary information*

- **S1.** The written procedure in full: the thirteen steps with their reasons, the 22-item report template, and the laboratory-constants block.
- **S2.** Session concordance. Sessions are named in the paper by curve, stage, and round; S2 gives for each the laboratory record identifiers and the server revision it ran on, together with the concordance between the specimen names used here (C1, Co1, CoC2 to CoC6, RuC1 to RuC5, Ru1, WB1, WB2) and the record names.
- **S3.** The expert's recorded rulings, verbatim and dated.
- **S4.** Results in full: the text and tables that Section 7 states in one paragraph per finding, with the $\chi^2$ like-for-like procedure.
- **S5.** The synthetic recovery benchmark: what the written procedure recovers against known truth, on the published engine.

- S6. Campaign 3, the final procedure on measured W/B4C: the design, the sessions, the scores against the tolerances, the order tolerance on two scales, and the roughnesses.

*** Not included in this preprint, provided upon request.**

## Data and code availability

The written procedure is published as its own record (Penkov et al., 2026b). It, the qualification scripts, the eight measured curves with their instrument metadata, the expert's reference projects, the session transcripts, the job files and fit reports, the figure scripts, the weighted-against-unweighted $\chi^2$ analysis of Section 7.5, and the tool-server revisions, each with the document that indexes it, are deposited at [http://github.com/OleksiyPenkov/xrr-fitting-skill-data](http://github.com/OleksiyPenkov/xrr-fitting-skill-data) and archived at doi:10.5281/zenodo.22851816. The six Ru/C curves and the expert's fits of them are those of a public data deposit (Penkov et al., 2026a).

## Declarations

**Funding.** This work received no external funding.

**Competing interests.** The authors declare no competing interests.

**Author contributions.** All three authors contributed to the work reported here.

**Use of AI.** The agent, the test sessions, and the drafting assistants are large language models, named at first use in Section 2.4. Every number in the text was checked by an author against the run record, and the authors are responsible for the content.